\documentstyle[12pt,epsfig]{article}
\author{L.J. Boya
\footnote{luisjo@posta.unizar.es}
\ M.A. Per
\footnote{maper@jazzfree.com}
\ A.J. Segu\'{\i}
\footnote{segui@posta.unizar.es}\\
Departamento de F\'{\i}sica Te\'orica. Facultad de Ciencias.\\
Universidad de Zaragoza. 50009-Zaragoza, Spain}
\title{Graphical and Kinematical Approach to Cosmological Horizons}
\begin{document}
\maketitle
\begin{abstract}
We study the apparition of event horizons in accelerated expanding cosmologies.
We give a graphical and analytical representation of the horizons using proper
distances to coordinate the events.
Our analysis is mainly kinematical. We show that, independently of the dynamical equations,
all the event horizons tend in the future infinity to a given expression depending
on the scale factor that we call \emph{asymptotic horizon}.
We also encounter a subclass of
accelerating models without horizons. When the ingoing null
geodesics do not change concavity
in its cosmic evolution we recover the de Sitter and
quintessence-Friedmann-Robertson-Walker models.
\end{abstract}
\vspace{2cm}
\vspace{.5cm}

\newpage
\tableofcontents
\vspace{3cm}
\section{ Introduction }
\label{sec:I}
The difficulty to make compatible string theory with accelerating
cosmologies has been studied in references \cite{r1} and \cite{r2}.
The de Sitter (dS) universe driven by a positive
cosmological term, as well as quintessence dominated universes,
posses event horizons; as a consequence two
physical systems can only be separated a finite distance, the
distance of the event horizon; then spatially
asymptotic states cannot exist and the S-matrix of interactions is not
well defined.

This fact contrasts with our understanding of string theory;
such a theory can be formulated only perturbatively where the initial
asymptotically free states (strings) interact giving rise to
a description of the final asymptotically free states in
terms of an S-matrix. This is the case for an asymptotically
flat background; if the geometry is asymptotically
anti-dS the asymptotic states are realized on the conformal
theory on the boundary, following Maldacena's conjecture \cite{Maldacena}
\footnote{For the dS case an analogous dictionary has been proposed
by Strominger \cite{r4} (see also \cite{r5}).}.
In general, gravitational backgrounds with horizons are not compatible
with a Hilbert space of infinite dimension like the Hilbert space
of string theory; in fact when gravity is present the number of degrees of
freedom that a system can support is proportional to the area of the system
according to the holographic principle \cite{'tHooft} \cite{susskind};
the event horizon, when present, is a natural place to project the relevant
degrees of freedom living on the bulk.

This theoretical difficulty contrasts with observations
\cite{Perlmutter} that, if confirmed, will prove that the observable
universe accelerates. Due to the attractive nature of the gravitational
force for ordinary matter these observations are compatible only with a
positive cosmological term, or with a sort of matter with negative
pressure termed quintessence, that can be associated to the vacuum energy
for an unstable scalar field evolving towards the minimum of its
potential; this sort of tachyonic instability can generate a vacuum
energy with negative pressure, a limit of which being the dS solution.

Assuming an equation of state for the matter of the type $p=w\rho$,
it can be shown that for an accelerating universe the range of $w$ is
$-1\le w < -1/3$. For a  Friedman Robertson Walker (FRW) universe
dominated by quintessence, $w$ is bounded by the previous values. $w=-1$
corresponds to a positive cosmological constant
($\rho=-p$) and the subsequent geometry is the dS one.
In \cite{r1} and \cite{r2} the formation of event horizons for the previous
values of $w$ has been studied showing that it is difficult to avoid them and, although
the observational data can be explained, it is not compatible with the 
string/M theory by the reasons commented above.

In spite that for positive cosmological constant $\Lambda$,
as well as for quintessence matter, the models of the universe both have event horizons,
they are different. In the case of quintessence the horizon
geometry is a cone with a singularity on the apex signaling the big bang
singularity, the topology being
${\mathbf{S}}^{2} \times {\mathbf{R}}^{+}(time)$. The dS horizon
geometry however is a cylinder, ${\mathbf{S}}^{2} \times {\mathbf{R}}(time)$, without
the singularity in $t=0$, and the distance to the horizon is constant and
equal to $\sqrt{3 \over \Lambda}$ so that the spatial limitation is constant
during the cosmological time. However, the proper distance to the horizon for a
quintessence dominated universe opens up with the cosmic time, being zero at the
big bang and growing linearly as the time elapses; the angle of the vertex of
the horizon cone depends on the value of $w$.

In the present research we study the presence of event horizons in a generic
way for metrics of the Robertson Walker (RW) type (spatially isotropic).
In order to locate them we
use a graphic representation that does not make use of the Penrose diagrams and
where the distances on the graphics are physical distances (proper or
coordinates); concretely we represent the successive light cones of a fiducial
observer comoving with the cosmic fluid, that we place in the origin, as well
as the world lines of galaxies
\footnote{The term galaxy is used freely illustrating
an emitter of light; we really refer to
a comoving dust element, or a geometric point.} without peculiar velocity,
that serve as fixed coordinate distance lines.
Another useful geometric
element is  the locus of comoving matter points that have a given, fixed
proper velocity with respect to the fiducial observer as a function
of cosmic time (\emph{iso-velocity lines}). This graphical representation is in
some sense complementary to the conformal mappings that give rise to the
Penrose diagrams, and has been used previously \cite{Ellis} with the aim of
clarifying the concept of particle horizons, putting
explicitly the presence of superluminal velocities  between different comoving
matter points of the fluid \cite{velocidades}. The velocity of any
object, \emph{locally} must be
smaller than $c$, the velocity of light in vacuo (locally, by the equivalence
principle, the theory of special relativity applies), but for separated
points, a non local measure, it is the theory of general relativity that
governs the phenomena.

For the explicit and exact location of the event horizon in a homogeneous expanding
cosmological model we need the explicit form of the scale factor $R(t)$
generally deduced from the dynamic equations. But only with a basic
information about the scale factor -without its explicit form- and by means of
the three kinematical tools previously commented -past light cones, world line
of galaxies and isovelocity lines-, can we determine the
existence or not of event horizons. We find that the decelerating cosmological
models do not have event horizon. However we also find a particular family of accelerated
cosmological models that neither have event horizon. Also we are able to
give an approximate expression that asymptotes the event horizon with
increasing cosmological time. The criteria that we will use to define the
location of event horizons are associated with the history of a photon emitted
by the galaxy towards the fiducial observer placed at the origin: when by its
travel the photon goes over cosmic regions with increasing proper velocities
with respect to the observer, then the emitted photon would never
arrive to the origin; in this case the galaxy is placed beyond the event
horizon. On the contrary, if in its path the photon enters regions where the
fluid velocity decreases (always with respect to the origin) the photon sooner
or later hits the origin. In the border we encounter the horizon (with respect
to the observer) that is placed where a emitted photon has a trajectory on
constant cosmic fluid proper velocities (with respect to the observer). We
emphasize that no reference is made to the velocity of the source, that is
arbitrary; also we do not prejudge the accelerated or decelerated character
of the universe.

The article is organized in five parts. We begin with the present
introduction. In the second part we discuss the graphical
representation method and we apply it to the quintessence and the dS cases.
In the third part we work in a generic cosmological setup and then we obtain
general results about event horizons. Finally we present our conclusions.
In a mathematical appendix we develop the argument of the main result.

\section{Graphical Emergence of Horizons}
\label{sec:II}
In this section we begin by introducing the expressions that would be used along
the article as well as the notation (that otherwise will be standard);
then we use them to study the event horizons for dS and quintessence
universes.

The RW geometry is characterized by the line element

\begin{equation}\label{2-1}
ds^{2}=-dt^{2}+R^{2}(t)\Big( {dr^{2} \over 1 - k \, r^{2} }
+r^{2}d\Omega_{2}^{2}\Big)
\end{equation}
where $d\Omega_{2}^{2}$ is the line element for the two dimensional unit sphere
${\mathbf{S}}^{2}$ and $k=\pm 1,0$ is the spatial curvature.
The coordinates are comoving with the cosmic fluid; in
particular $t$ is the cosmic time. The velocity of light is $c=1$.
Due to the isotropy of the geometry (in
accordance with the cosmological principle) the origin is a generic point and
we can fix one direction (fixing the polar angles) without loss of generality;
the metric takes then the form

\begin{equation}\label{2-2}
ds^{2}=-dt^{2}+R^{2}(t) {dr^{2} \over 1 - k \, r^{2}  } \ ,
\end{equation}
where each point of the plane $t-r$ must be considered as a sphere of
radius $R(t)r$. Instead of using the radial coordinate $r$, the events
can be coordinated using the proper distance to the origin $r=0$ for a
given time $t$; that is taking $dt=0$ in (\ref{2-2})
$D(t,r)=R(t)\int_{0}^{r} {dr' \over \sqrt{1 - k r'^{2} } }\ $;
differentiating we have

\begin{equation}\label{2-3}
{dr \over \sqrt{1 - k r^{2} } }={dD \over R}-{D \over R^{2} } \ {\dot R} \ dt \ ;
\end{equation}
substituting in (\ref{2-2}) gives

\begin{equation}\label{2-4}
ds^{2}=-dt^{2}+ \Big( dD- D \  { {\dot R} \over R} \ dt \Big) ^{2}.
\end{equation}
Each point in the plane $t-D$ continues to be a sphere of radius $D$.

Let us remember the three ingredients we will use to fix the
event horizons:
\begin {enumerate}
\item
The locus of the ingoing null geodesics (NGs) expressed in proper distance.
Concretely the \textbf{past light-cone} of a fiducial observer that we will
place at the origin $r=0$. 
\item
The \textbf{world line of test galaxies}
without peculiar velocity and its proper distance to the observer (its
coordinate distance $r$ will be constant). 
\item
The proper distance to the
origin, as a function of cosmic time, for the galaxies that move with a fixed
proper velocity with respect to the observer: \textbf{isovelocity lines}. This last
concept will be developed in the next section. 
\end{enumerate}

In order to determine the NGs and consequently the causal structure
of the geometry we put $ds^{2}=0$ in (\ref{2-4}), yielding
\begin{equation}\label{2-5}
{dD \over dt}= D { {\dot R} \over R} \pm c \ ;
\end{equation}
here we put explicitly $c$, the light velocity.
$D$ measures the proper distance to the
origin for a photon emitted by a galaxy with a given initial coordinates. There
are two contributions to the velocity of the photon with respect to the
observer in the origin: $\pm c$ is the local velocity of light with respect to
the geometrical background; locally, as a consequence of the equivalence
principle, the geometry is minkowskian and are valid the postulates of the
theory of special relativity; the photon moves with constant velocity $c$
respect to the surrounding matter, $c$ being an upper bound; the two signs of
the second term of (\ref{2-5}) corresponds to the two possible directions
away ($+$) or towards ($-$) the origin. The term $H(t)D$
($H={ {\dot R} \over R}$ is the Hubble constant) represents the photon dragging
as a consequence of the variation of the geometry with time; the space-time
geometry is not flat
and special relativity does not apply, but the general one. (\ref{2-5})
shows that the photon velocity with respect to the origin can take arbitrary
values
\footnote{In (\ref{2-5}) we recover the newtonian sum of velocities; this
is a signal that the motion of comoving (without
proper velocities) galaxies is due to the expansion of the space, not
a motion \emph{on the space}. See \cite{velocidades}.};
the arbitrariness is encoded in the functional form of the scale factor
$R(t)$ that depends on the way the geometry is governed by its coupling with
matter (ordinary or vacuum energy).

The relation (\ref{2-5}) is purely geometric. To obtain the causal structure
of the space time explicitly we need to know the nature of the matter
that couples to the geometry. We suppose that the substratum is an ideal fluid
characterized by the equation of state $p=w \rho$. We restrict ourselves
to the  spatially flat case $k=0$ that seems to fit well with the
observations \cite{Perlmutter}. Then the Einstein equations
in the FRW form are
\begin{equation}\label{2-6}
H^{2}={8 \pi G \rho \over 3},
\end {equation}
\begin{equation}\label{2-7}
{\dot \rho}+ 3 H (\rho+p)=0.
\end{equation}
The equation of continuity (\ref{2-7}) is independent of the
constant curvature of the spatial sections.
Using the equation of state we can integrate it
and obtain the dependence of the density with the scale factor;
for $\omega$ constant we obtain
\begin{equation}\label{2-8}
\rho \propto {1 \over R^{3(1+ \omega)}}.
\end{equation}
The two FRW equations can be recast, integrating (\ref{2-6}) with the aid of
(\ref{2-8}) and differentiating (\ref{2-6}) with respect to $t$, 
with the following result
\begin{equation}\label{2-9}
R(t)=R(t_{0}) \Big({t \over t_{0} } \Big)^{2 \over 3(1+\omega)} \ ,
\end{equation}
\begin{equation}\label{2-10}
{ {\ddot R} \over R} =-{4 \pi G \over 3} (1+3 \omega) \rho \ .
\end{equation}
The Hubble constant is
\begin{equation}\label{2-11}
H(t)={2 \over 3(1+ \omega)\ t}.
\end{equation}
We observe that for $\omega=-1$ the power law dependence of the scale factor
(\ref{2-9}) breaks down; in fact for $p=- \rho$ the fluid is equivalent to the
presence of a positive cosmological constant and the appropriate model is
the dS universe with an exponential expansion.
The equation (\ref{2-10}) proves that for $w<-{1 \over 3}$ the expansion
($ {\dot R}>0$) is accelerating ($ {\ddot R}>0$).
The range $-1< \omega<-{1 \over 3}$ characterizes the quintessence.

To obtain the possible cosmological horizons the usual way to 
proceed is the following. Given the matter that fills the universe,
the $\omega$ that can depend on the cosmic time, we solve the equations
(\ref{2-6}) and (\ref{2-10}), and obtain the explicit solution for the
scale factor $R(t)$. Using the NGs of the metric (\ref{2-2}) we determine
the space-time regions that will never be observed and
the proper distance to the event horizon as a function of $t$:
\begin{equation} \label{2-100}
D_{Hor}(t)=R(t) r_{Hor}(t)=R(t) \int_{t}^{\infty} {dt' \over R(t')}.
\end{equation}

In order to determine the cosmological event horizons we use
a slightly different method.
First, we obtain from the metric (given the explicit function $R(t)$)
the family of ingoing NGs that reach $r=0$ when $t=t_{0}$, i.e. the family
of past light cones for an observer at $r=0$. We represent
graphically this family of ingoing NGs using proper distances to coordinate them;
then we observe an accumulation of past light cones
towards a limiting one that will trace the event horizon. Beyond this limit the ingoing
NG will never reach $r=0$ and then those space-time region would never be
observed. In the following subsections we apply this procedure to the accelerating models,
FRW dominated for quintessence and dS, for the flat spatial case.

\begin{figure}[!hbt]
\begin{center}
\includegraphics[width=14cm]{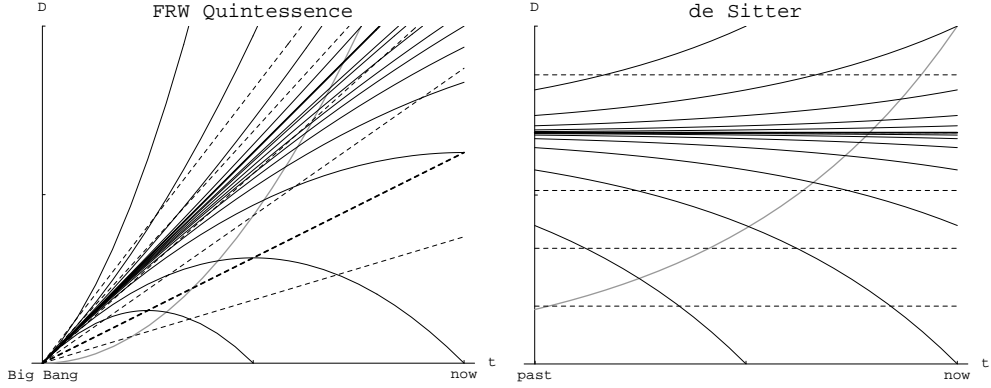}
\end{center}
\caption{{\small Graphical elements for the location of the event horizons.
Ingoing NGs (past light cones), world line of a galaxy (gray line)
and isovelocity lines (dotted lines) including the Hubble distance (heavy dotted line).
We observe that, for the dS and quintessence cases, the NGs don't explore all
the space-time events. They accumulate on a NG limit (thick line) that corresponds to
the event horizon.
We observe that the galaxy crosses the event
horizon at a finite time but the observation time $t_{0}$ diverges,
as for the black hole case.
}}
\label{figura1}
\end{figure}

\subsection{The case of quintessence}
\label{subsec:II-1}

We integrate the NG differential equation (\ref{2-5})
for the quintessence dominated universe. Using the value of
the Hubble constant (\ref{2-11}) we can write
\begin{equation}\label{2-12}
D(t)=\pm {3(1+\omega) \over 1+3 \omega} t+ K t^{2 \over 3(1+\omega)} \ ,
\end{equation}
where $K$ is an integration constant; forcing the light trajectory to
be at a  proper distance $D_{0}$ from the origin when the cosmic time
is $t_{0}$ results in
\begin{equation}\label{2-13}
D(t)=\pm {3(1+ \omega) \over 1+3 \omega}\, t
    + \Big[D_{0} \mp {3(1+ \omega) \over 1+3 \omega} t_{0} \Big]
         \Big({t \over t_{0} } \Big)^{2 \over 3(1+ \omega)} \ .
\end{equation}
In this way we construct the two families of NGs, one with growing $D$
and the other with decreasing $D$. $D(t)$ according to (\ref{2-13}) is the light cone
for each point of the manifold $(D_{0},t_{0})$. The families (\ref{2-13})
determine the causal structure of the universe.

The past light cone for the origin $r=0 (D=0)$ at time $t_{0}$ is given by
\begin{equation}\label{2-14}
D(t)= {3(1+ \omega) \over 1+3 \omega}\, t_{0}
    \Big[  \Big({t \over t_{0} } \Big) ^{2 \over 3(1+ \omega )} - {t \over t_{0}}\Big] \ .
\end{equation}
Taking the derivative of the previous expression with respect to the cosmic
time gives us the slope of the  past light cone
\begin{equation}\label{2-15}
\dot{D}(t)= {3(1+ \omega) \over 1+3 \omega} \,
    \Big[{2 \over 3(1+ \omega)}\,
    \Big({t \over t_{0} } \Big)^{-{1+3 \omega \over 3(1+ \omega)}}
        -1 \Big] \ ;
\end{equation}
we see that $\dot{D}(t)$ is a decreasing function of $t$,
its maximum value occurs for $t=0$;
because $-{3(1+ \omega) \over 1+3 \omega}$ is positive for $-1< \omega < -{1 \over 3}$
the slope of the light cone at the initial time is finite and is
\begin{equation}\label{2-16}
\dot{D}(t=0,\forall \, t_{0})={-3(1+\omega) \over 1+3\omega}>0 \ .
\end{equation}
This means that the events that can influence the observer placed at the origin
are limited to the cone with apex in the big bang and with an angle
$\tan^{-1} [-{3(1+ \omega) \over 1+3 \omega}]$.

To determine the event horizon we must compare the past light cone of the origin
with the world lines of the galaxies. The recession of a galaxy that in $t_{0}$
is placed at $D_{0}$ is described by
\begin{equation}\label{2-17}
D(t)=D_{0} \Big({t \over t_{0}}\Big)^{2 \over 3(1+\omega)} \ .
\end{equation}
$\omega=-{1 \over 3}$ separates two regimes with different concavity for $D(t)$
reflecting the accelerated/decelerated behavior of matter. The derivative
of (\ref{2-17}) with $t$ is ${\dot D}(t)\propto {t^{-{ 1+3 \omega \over 3(1+ \omega)}}}$;
this derivative at the origin is zero for the quintessence but diverges for
$\omega>-{1 \over 3}$; the big bang for the quintessence is not big nor bang,
however the quintessence density diverges for $t\rightarrow 0$ as $t^{-2}$.
Now, it is
clear that every pair of galaxies has been causally connected in the beginning
of the expansion; also it is clear that the same pair of galaxies would be
causally disconnected when one of them crosses the event horizon of the
other. The horizon grows linearly with the cosmic time according to
\begin{equation}\label{2-18}
D_{Hor}(t)=-{3(1+\omega) \over 1+3 \omega}\, t \ .
\end{equation}

For ordinary matter $-1/3 \leq \omega \leq 1$,  $D_{Hor}=\infty$. We can use
(\ref{2-14}) to study the future light cone; taking $t_{0}=0$ and changing sign
we obtain the spatial extension of the future light cone from the big bang
i.e. the particle horizon, so that for ordinary matter,
\begin{equation}\label{2-190}
D_{PH}(t)={3(1+\omega) \over 1+3 \omega}\, t \ ,
\end{equation}
and for quintessence ($-1< \omega <-{1\over 3}$), $D_{PH}=\infty$;
particle and event horizons play a dual role (see Table \ref{tabla1}).

\begin{figure}[!hbt]
\begin{center}
\includegraphics[width=10cm]{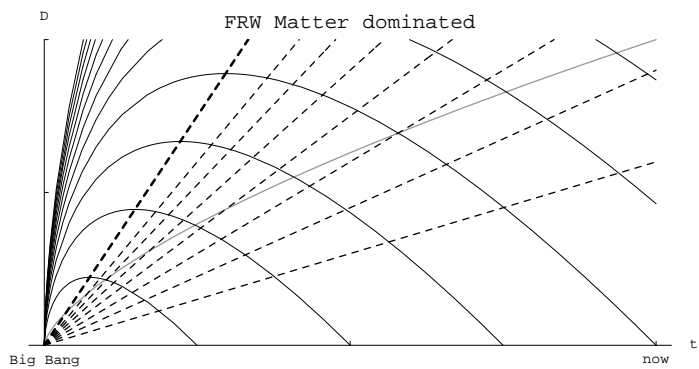}
\end{center}
\caption{{\small Causal structure for a decelerated model with $\omega=0$.
We observe that the photons emitted by a galaxy (grey line) cross the
isovelocity lines (dashed lines) always on the decreasing recession velocity sense.
There are not real
accumulation of past light cones indicating the absence of event horizons for this case;
for every space-time event $(D,t)$ there exists an ingoing NG reaching $D=0$.}}
\label{figura2}
\end{figure}

In Fig. \ref{figura1} we show the world line for an arbitrary galaxy
together with the past light cones of an observer placed at the origin; in this
way we visualize the event horizon which is also shown.
Beyond this limit all the ingoing NGs never reach $D=0$ and
so we will never observe at $D=0$ the events covered by this NG.
We have also depicted
the same functions for ordinary matter ($\omega>-{1 \over 3}$) in
Fig. \ref{figura2}. Here the expansion is decelerated
and as a consequence the slope of the  past light cone at the initial
time diverges; we need an infinite amount of kinetic energy to overcome the
(also infinite) amount of gravitational attraction; the world line of a generic galaxy,
having now contrary concavity to the quintessence, will intercept unavoidably
the past light cone of the observer.

\subsection{The case of $\Lambda>0$.}
\label{subsec:II-2}

The same reasoning can be done for the case of a
positive cosmological constant ($\Lambda>0$)
dominated universe. The dS model corresponds to an equation of
state $p=-\rho$ ($\omega=-1$). For the FRW equations (\ref{2-6}) and (\ref{2-7})
it follows than the scale factor goes exponentially with time
$R(t)=R_{0} e^{\pm \sqrt { {\Lambda \over 3}} t }$, and the continuity equation
shows manifestly the constancy of the vacuum energy
$\rho_{\Lambda}= { \Lambda \over 8 \pi G}$.  $R_{0}$ is an arbitrary scale factor
for $t=0$ because the model, possessing an exponential expansion, is self similar.
The Hubble constant is in this case truly constant
$H=\pm \sqrt{ {\Lambda \over 3} }$ and the model is in agreement with
the perfect cosmological principle.

To obtain the causal structure we need to know the past light cone of
a generic observer as well as the world line of the galaxies
\footnote {The galaxies are test particles or geometric tracers; we suppose
that all the energetic contribution is due to the cosmological term.}.
Again we make use of the proper metric (\ref{2-4}) and integrating the
NGs we arrive to
\begin{equation}\label{2-19}
D(t)=e^{-H t_{0}} \Big( D_{0} \mp {1 \over H }  \Big) e^{H t} \pm {1 \over H} \ .
\end{equation}
The previous expression is the light cone for an event placed at the point
$(t_{0},D_{0})$; taking the origin $D_{0}=0$ $(r_{0}=0)$ at $t=t_{0}$ as our fiducial
observer we see that such an event can be influenced by the events inside the region
\begin{equation}\label{2-20}
D(t)= \sqrt{ {3 \over \Lambda}}  \Big( 1- e^{\sqrt{{\Lambda \over 3}} (t-t_{0})}  \Big) \ ,
\end{equation}
where obviously $t<t_{0}$. One observes that the past light cone never extends to
distances greater than $\sqrt{ {3 \over \Lambda}}$. The world lines of
galaxies have the form
\begin{equation}\label{2-21}
D(t)=D_{0} e^{ \sqrt{ { \Lambda \over 3}} (t-t_{0}) },
\end{equation}
and in a finite cosmic time overtake proper distances with respect to the origin
higher than $\sqrt{ {3 \over \Lambda}}$. The proper distance
$\sqrt{ {3 \over \Lambda}}$ is the event horizon for the observer placed at
the origin and consequently for any observer. The geometry of the
event horizon for the dS universe is cylindrical in front to
the conical geometry for the quintessence dominated universe.
In Fig. \ref{figura1}
we show the dS model with its causal structure;
we plot the successive past light cones from $D=0$.
A typical galaxy world line crossing the isovelocity lines has also been included.

\begin{table}[!htb]
\begin{tabular}{|c|c|c|c|c|c|c|}
\hline
  & & & & &$\ _\textrm{light} \ $ & \\
  & Event & Particle & Appar.

  &Hubble & cone& galaxy \\
  & Horizon & Horizon & Horizon & Distance & $\dot D_{n}$ & $\dot D_{r}$ \\
 Universe: \quad &$D_{Hor}$ &$D_{PH}$ & $D_{n}^{Max}$ & $v=c$ & $ ^{t\rightarrow 0} $
 & $^{t\rightarrow 0}$ \\
\hline \hline
 & & & & & & \\
   $\ddot{R}<0$&
   $\infty$ &
   $\frac{3(1+\omega)}{1+3\omega} \ t$ &
   $\ t$&
   $\frac{3(1+\omega)}{2} \ t$ &
   $\infty$ &
   $0$ \\
 $ ^{\omega \in (-\frac{1}{3},1]}$& & & & & & \\
 \hline
   & & & & & & \\
   $\ddot{R}=0$ &
   $\infty$ &
   $\infty$ &
   $ t$ &
   $ t$ &
   $\infty$ &
   $\frac{D_{r0}}{t_{0}}$ \\
  $ ^{\omega = -\frac{1}{3}}$& & & & & & \\
 \hline
   & & & & & & \\
   $\ddot{R}>0$ &
   $\frac{-3(1+\omega)}{1+3\omega} \ t$ &
   $\infty$ &
   $\ t$&
   $\frac{3(1+\omega)}{2} \ t$ &
   $\ \frac{-3(1+\omega)}{1+3\omega}\ $ &
   $\infty$ \\
  $ ^{\omega \in (-1,-\frac{1}{3})}$& & & & & & \\
 \hline
   & & & & & & \\
   $dS$ &
   $\sqrt{\frac{3}{\Lambda}}$ &
   $\infty$ &
   $\sqrt{\frac{3}{\Lambda}}$ &
   $\sqrt{\frac{3}{\Lambda}}$&
   - & - \\
 $ ^{\omega =-1}$& & & & & & \\
 \hline
\end{tabular}
\caption{{\small
\emph{Event} Horizon, \emph{Particle} Horizon,
\emph{Apparent} Horizon and the Hubble distance for FRW and dS flat cosmological
models. We show also the behavior of the slope of the
galaxy world line and past light cone near the Big Bang. }}

\label{tabla1}
\end{table}

\section{General results. The Asymptotic Horizon}
\label{sec:III}
In this part we give a general criterion to place an event
(the emission of a photon by a galaxy) beyond or below the possible event horizon
of a generic observer. We analyze the recession velocity of the regions traversed by the
photon emitted by the galaxy towards the observer at the origin. If this cosmic
velocity increases we affirm that the photon never reaches the origin and the
photon emitter galaxy is hidden by the horizon; because the local velocity
with respect to the cosmic fluid is bounded by $1$, the same result happens for
any other carrier of information leaving the galaxy. On the contrary, if the
photon's path goes over cosmic regions with decreasing proper velocities with respect
to the observer, the light sooner or later will overtake the origin; the emitter
is causally connected with the origin. The event horizon is the border between
the two previous situations.

\subsection{Generic Cosmological Models}

We are dealing with cosmologies described by a Robertson-Walker geometry characterized
by the scale factor $R(t)$ and the curvature that we take null $(k=0)$. We can
obtain valuable information by kinematic as well as geometric arguments deduced for
the metric independently of the dynamics, that is, the particular way the geometry
couples to matter via the Einstein-Friedman equations. To this end we remember some
well known expressions that will fix notation.
\begin{description}
\item[Galaxies-Particles-Tracers]The elements of the cosmic fluid are comoving and so
the spatial coordinates $r$ are constant. All the spatial points being equivalent,
we take the origin as the reference point. The proper distance of a particle with
coordinate $r$, that at time $t_{0}$ is at proper distance $D_{0}$, is
$D_{r}(t)=r R(t)= D_{0} R(t)/R(t_{0})$. From this we obtain the Hubble law which can
be rewritten and reinterpreted as an \emph{isovelocity law}
\begin{equation}\label{3-1}
D_{v}(t) =v {R(t) \over \dot R(t)} \ ,
\end{equation}
which is the proper distance to the origin of fluid points, with constant
proper velocity $v$, as a function of time.

Using (\ref{3-1}) we can identify the space-time points where the recession velocity of the
galaxies with respect to the origin is $1$; locally the special relativity
governs the phenomena and the light velocity with respect to the galaxies is also $1$;
consequently, the photons emitted by such galaxies towards us have zero velocity
with respect to the origin; this fact translates in the development of a maximum
for the NGs in such points. The locus where this maxima appears for $k=0$
\footnote{For a very interesting discussion about the implications
of $k=\pm 1$ and the kinematical/dynamical character of
the cosmological lens effect see \cite{Ellis2}} defines the
\emph{apparent horizon} and matches the isovelocity line for $v=1$; this defines the
Hubble distance $D_{Hub}= {R \over \dot R}$. The apparent horizon has been proposed
as a holographic screen to realize the holographic principle in cosmology
\cite{Fishlersusskind} \cite{bousso}
and shares many properties with the black hole horizon, as the locus of trapped surfaces.

\item[Null Geodesics] The evolution of the NGs that we have previously
identified from the metric (\ref{2-5}) can also be obtained imposing that the velocity of
the NG with respect to the galaxies that it crosses equals $\pm c$; using
the Hubble law (\ref{3-1}) we obtain the velocity of the NGs with respect to us:
\begin{equation}\label{3-2}
\dot D_{n}(t)=\dot D_{r}(t) \pm 1=D_{n}(t) {\dot R(t) \over R(t)} \pm 1 \ ,
\end{equation}
where $D_{n}(t)$ is the proper distance to the NG.
The ingoing NGs that meet the origin at $t_{0}$ can also be obtained by
direct integration of the metric (\ref{2-2}),
\begin{equation}\label{3-3}
D_{n}(r=0,t_{0})(t)=R(t) \int_{t}^{t_{0}} {dt' \over R(t')} \ ,
\end{equation}
and describes the universe we see at $t_{0}$; however we need
the function $R(t)$ given by the dynamics to obtain its explicit form.
The NG that traverses
the point $(D_{0},t_{0})$ can be obtained, using the homogeneity hypothesis,
by summing the distance between the galaxy and the origin with the distance
between the photon and the galaxy
\begin{equation}\label{3-4}
D_{n}(D_{0},t_{0}) (t)=D_{0} {R(t) \over R(t_{0}) } +D_{n}(0,t_{0}) (t) \ .
\end{equation}
It is easy to verify that (\ref{3-4}) satisfies the geodesic condition (\ref{3-2}).

\item[Event horizon] In the previous section we have obtained the event horizon
identifying the space-time regions that never can be reached by an ingoing null
geodesic; the term \emph{never} implicitly means
to take the limit $t_{0} \rightarrow \infty$;
then we need to know the behavior of the function $R(t)$ for all time
and consequently the dynamics. The expression for the event horizon in terms
of the scale factor is given by (\ref{2-100}).
This function represents the temporal evolution of the ingoing NG frontier
between photons observed (reaching $D=0$)
and the unobserved ones $(D \rightarrow \infty)$.
\end{description}

\subsection{The Asymptotic Horizon}
Without recourse to the dynamics and therefore valid for a variety of cosmological
models we can study the points where the NGs change its concavity by differentiating
(\ref{3-2}), with the result
\begin{equation}\label{3-6}
\ddot D_{n}={ D_{n} \ddot R \mp \dot R \over R}=0  \Rightarrow
D_{AsH}=\pm {\dot R \over \ddot R} \ ,
\end{equation}
where the positive sign for $D_{AsH}$ corresponds to ingoing
geodesics; $D_{AsH}$  refers to the
\emph{asymptotic horizon}. We will show that the event horizons of any cosmological model,
will converge, in the future infinity, with the asymptotic horizon previously defined.
The rigorous demonstration is left to the appendix, and here
we merely argue its plausibility.

If the cosmological model is of the FRW (quintessence) or dS type it is easy to prove
that the asymptotic horizon and the event horizon coincides for all time. For the FRW models,
the scale factor is given by (\ref{2-9}); then the value of $D_{AsH}$ as defined in
(\ref{3-6}) is $D_{AsH}=- { 3(1+\omega) \over 1+3 \omega }t$; this matches the value of
the event horizon (\ref{2-18}) for all time. For the case of a cosmological constant
driven universe the scale factor goes as $R=R_{0} \, e^{t \sqrt{\Lambda / 3} } $;
the asymptotic horizon is placed at $D_{AsH}= \sqrt{ 3/ \Lambda} $ which is the distance
to the event horizon independent of the cosmic time.

\begin{figure}[!hbt]
\begin{center}
\includegraphics[width=14cm]{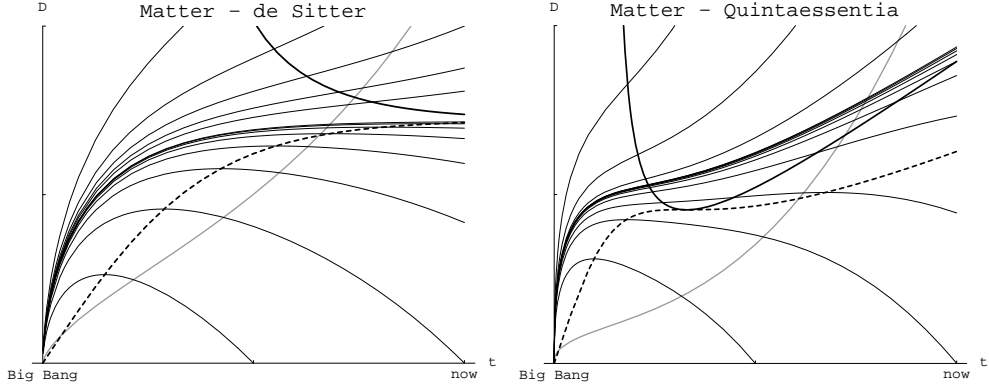}
\end{center}
\caption{{\small The convergence between the event horizon and the asymptotic horizon is
shown for the two cases discussed in text. On the left
a universe with dust ($\omega=0$) and a positive cosmological constant is represented.
This case corresponds to the inside-outside evolution of the NGs
(see the text). The figure on the right,
outside-inside evolution, corresponds to a mixture of quintessence ($\omega=-3/4$) and
matter ($\omega=2/3$). We see also the concavity change for the NGs when
crossing the asymptotic horizon.}}
\label{figura3}
\end{figure}

In any other expanding accelerating model a part of the NGs crosses the asymptotic horizon
at least once; let us follow them for the last time they traverse the asymptotic horizon.
This can be done in two different ways, namely:
\begin{itemize}
\item \emph{inside-outside}. This is the usual case (see Fig. \ref{figura3} left).
The relevant part
of the bundle of NGs comes from the inside region
of the asymptotic horizon $(D_{n}<D_{AsH},\ddot D_{n}<0)$ and part of them goes outside
$(D_{n}>D_{AsH},\ddot D_{n}>0)$. All the NGs outside the asymptotic horizon
increases the distance to the origin without limit. The event horizon will be traced
by the NG that approaches the asymptotic horizon from inside but without crossing it;
this will be the NG always below the asymptotic horizon without maximum. In this case
$D_{Hor}<D_{AsH}$ and $\ddot D_{Hor}<0$.

\item \emph{outside-inside}.
The part of the bundle of NGs we are interested (see Fig. \ref{figura3} right),
advances as the cosmic time
elapses in the outside part of the asymptotic horizon $(D_{n}>D_{AsH}, \ \ddot D_{n}>0)$;
part of the bundle will traverse it and enters the inside part
$(D_{n}<D_{AsH},\ddot D_{n}<0)$.
Then they reach a maximum distance to the origin and after this, approaches $D=0$ until
they meet the origin and are observed. The horizon is defined by the NG that
approaches progressively the asymptotic horizon
from the outside part but without traversing it. In that case
$D_{Hor}>D_{AsH}$ and $\ddot D_{Hor}>0$.
\end{itemize}

The two situations previously described are the more general ones. The proof
is left to the mathematical appendix.

There is a relation between the deceleration parameter $q=-{ \ddot R R \over \dot R^{2} }$
and $D_{AsH}$ namely
\begin{equation}\label{3-9}
q=-{ D_{Hub} \over D_{AsH} } \ ;
\end{equation}
it is easy to show that
\begin{equation}\label{3-10}
\dot D_{Hub}=1+q \ ;
\end{equation}
now, if $\ddot R>0$ ($D_{AsH}>0$) then $q<0$; if $D_{Hub}<D_{AsH}$ then $q \in (-1,0)$ and
using (\ref{3-10}) it follows that $\dot D_{Hub}>0$; more concretely
$\dot D_{Hub} \in (0,1)$.

Let us discuss when $D_{Hub}<D_{AsH}$ and the previous arguments applies. In general
the ingoing geodesics have maxima that force them to converge until they are observed.
The maxima are localized on $D_{Hub}$ and being maxima $\ddot D_{n}<0$. In the
accelerating models, because at $D_{AsH}$ there is a change of concavity for the
NGs, $D_{Hub}$ must be below  the asymptotic horizon.
Nevertheless, due to changes of the sign of the acceleration it is possible
that $D_{Hub}>D_{AsH}$; this would causes the existence of minima in the NGs.
It is easy to prove that the presence of minima in the NGs implies the violation of
the dominant energy condition \cite{HE}. For an equation of state of the type
$p=\omega \rho$, we have with
the use of the Friedman equations (\ref{2-6}) and (\ref{2-10}),
\begin{equation}\label{3-100}
q=-{\ddot{R} \over R} {1 \over H^{2}}={ 1+3 \omega \over 2} \ .
\end{equation}
If the deceleration parameter is smaller than $-1$ (which is the
condition for the presence of minima in the NGs)
then $\omega <-1$, violating the dominant energy condition.
So, for an expanding model, we will consider the realistic assumption $\dot D_{Hub}>0$.

\subsection{No-horizon Universes} We can now prove using kinematical
arguments the following statement: \emph{ Universes with a final expansive decelerating
epoch do not have event horizons}. If $\ddot{R}<0$ then $\dot{R}$ is a decreasing
function, so $\lim_{t \rightarrow \infty} \dot{R}=a<\infty $; then for later times
$t>t_{c}$  there exist $\tilde{a}>a$ so that the scale factor has
an upper bound $\tilde{a} \ t > R(t)$; the distance to the horizon diverges
\begin{equation} \label{3-3-2}
D_{Hor}(t)= R(t) \int^{\infty}_{t} \frac{ dt'}{R(t')} \ > \
R(t) \int^{\infty}_{t_{c}} \frac{ dt'}{\tilde{a} \ t'} \ \rightarrow \ \infty \ .
\end{equation}

We have shown that the necessary condition for the
emergence of cosmological horizons in expansive universes is that they have to
be accelerated at their final epoch. Then $D_{AsH}>0$ and so the emergence of
event horizons imply the concavity change of some of the NGs \footnote{Unless
the asymptotic horizon coincides with the event horizon for all times, which is the
case of quintessence and dS models.}
(the NGs that intersect $D_{AsH}$). But not all the accelerating universes have event
horizon; if $\dot{R}(t)\geq 0$ and $\lim_{t\rightarrow \infty}\ddot{R}=0^{+}$, although
$\ddot{R}(t)>0$ for all $t$, \emph{all} the NGs intersect $D_{AsH}$ from outside to inside
and finally reach $r=0$.
The demonstration of this last statement is identical to the previous one. Again,
we can bound from above the scale factor with a linear function of $t$
and proceed as in (\ref{3-3-2}).
It is easy to prove also that this family of universes verify
$\lim_{t\rightarrow \infty}\dot{D}_{Hub}=1^{-}$ and
that $\lim_{t\rightarrow \infty}\dot{D}_{AsH}=\infty$ (see (\ref{5-C}) in
the Appendix).
There is an infinity of functions obeying this condition and it is possible to
study (with the dynamical equations) the relation between the equation of state,
that is the value of $\omega (t)$, and the functional form of the scale factor
(work in progress).

\begin{figure}[!hbt]
\begin{center}
\includegraphics[width=8cm]{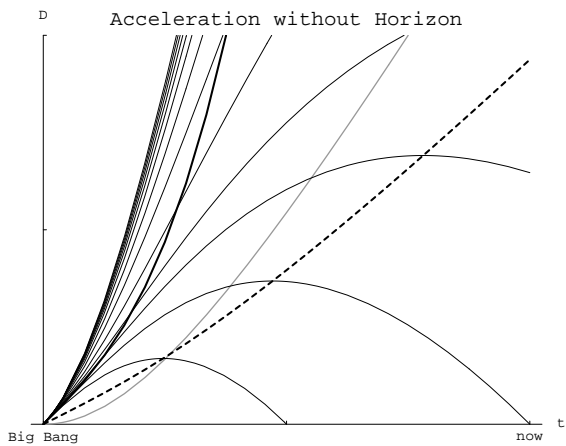}
\end{center}
\caption{{\small For an
accelerating model with acceleration tending to zero there is not
accumulation of past light cones and therefore all the space-time events are
traversed by them signaling the absence of event horizons.
We also represent the Asymptotic Horizon (thick line);
we can check its divergent slope and the concavity change of the ingoing NGs there.}}

\label{figura4}
\end{figure}

\subsection{Kinematical deduction of the event horizons for FRW and dS universes}
Let us follow the path for a photon emitted by a galaxy in order to see if it can reach
the origin and can be observed or not. If the velocity of the galaxy is bigger
than $1$ the photon will travel to us with negative
velocity and will go away. If the universe
is accelerating, the photon in its travel traverses regions with higher fluid velocity
and its relative proper velocity with respect to the origin grows.
The first statement is true;
the second false, because the photon evolves also in the temporal direction, and in this
direction the velocity of the galaxies always decreases;
this can be seen differentiating the
Hubble law $v(t)=D/D_{Hub}$ for fixed $D$
\begin{equation}\label{3-8}
{ \partial v \over \partial t}=-D { \dot D_{Hub} \over D_{Hub}^{2}}<0,
\end{equation}
because $\dot D_{Hub}>0$; in fact $\dot D_{Hub} \in (0,1)$  in the accelerating epochs
\footnote{This arguments can be used to clarify the issues of the \emph{chained
galaxy} paradox \cite{encadenada}.}.

We see that the final history of the photon depends
on the way the galaxy recession velocity field
is distributed. This is described by the isovelocity bundle
(\ref{3-1}); therefore, if the evolution of the photon traverses the isovelocity
lines towards growing values then this photon goes away from the origin with increasing
relative velocity being \emph{probable} it will never return to $r=0$; and vice versa.
In the first case $\ddot D_{n}>0$, in the other case $\ddot D_{n}<0$.

The \emph{probable} horizon will be placed where $\ddot D_{n}=0$, and we have seen that
such points are localized by the function $D_{AsH}(t)$ (\ref{3-6}). What happens in
$D_{AsH}$ is that a NG ceases to traverse the isovelocity lines in the
increasing (decreasing) sense, during some instants is parallel to one of them and finishes
traversing the bundle in the decreasing (increasing) direction. The positive sign for
$D_{AsH}$  and consequently the existence of this \emph{probable} horizon implies an
accelerating universe\footnote{We do not consider the decelerated contracting universes
where also $D_{AsH}>0$.}.
This horizon is \emph{probable} and not \emph{sure}; it can happens that a NG exploring
regions of growing recession velocities, and that we suspect to be unobservable, in a given
time crosses the line $D_{AsH}$ and by changing its concavity
begins to travel regions characterized
by decreasing recession velocities respect to the origin arriving at the end to $r=0$.
To avoid this situation and in this way to confirm that the \emph{probable} horizon is the
\emph{sure} horizon, we must be certain than each of the NGs cannot change concavity.
This does not means that all of them have the
same concavity because in this case $D_{AsH}<0$.
The situation we consider corresponds to two families of NGs with different
concavities separated by a lineal NG that forms the border between them and that
matches $D_{AsH}$.

What cosmological model verifies than $D_{AsH}$ is the locus of an ingoing NG?
If this condition is satisfied we can affirm that $D_{AsH}$ is the event horizon of such
models of the universe. If $D_{AsH}$ is an ingoing NG by no means will be
traversed by any other one (the NGs form a bundle that only at the singular
points like the Big Bang can meet). As a consequence the rest of the NGs
have a definite and invariable concavity; $D_{AsH}$ will be then linear $D_{AsH}=at+b$.
We appreciate two possibilities:
\begin{itemize}
\item Case $a\neq 0$. Introducing $D_{AsH}$ in the geodesic equation (\ref{3-2}) we obtain
\begin{equation}\label{3-11}
{ a+1 \over at+b}={ \dot R(t) \over R(t)} \quad \Rightarrow \quad
R(t)=R_{0} t^{1+{ 1 \over a} } \ ,
\end{equation}
where we impose $R(0)=0 \ (b=0)$. We have obtained a FRW solution with the temporal
power given in terms of the slope $a$ of the event horizon instead of the parameter of the
equation of state $\omega =p/\rho $. The slope of the event horizon is then
\begin{equation}\label{3-12}
a=-{ 3(1+ \omega) \over 1+3 \omega} \ .
\end{equation}
\item Case $a=0 \Rightarrow b \neq 0$. Substituting the value of $D_{AsH}$ into (\ref{3-2})
it follows
\begin{equation}\label{3-13}
{1 \over b} ={\dot R(t) \over R(t) } \quad \Rightarrow \quad
R(t)= R_{0} e^{t/b}.
\end{equation}
In this case we obtain a dS universe with $b$, the distance to the horizon, related
with the cosmological constant $\Lambda$ according to
\begin{equation}\label{3-14}
b=\sqrt{ {3 \over \Lambda } } \ .
\end{equation}
\end{itemize}

\section{Discussion and conclusions}
\label{section:IV}
The coordinates used in this work in order to describe the causal structure of RW
cosmological scenarios are not the usual ones; we have used proper distances to comoving
observers as coordinates to represent the light ray paths and we have been able
to extract the event horizons of the models studied in this way. The standard method
used to obtain the causal structure is to bring the infinities at a
finite coordinate value making use of a conformal transformation of the metric;
the light cones in the new metric are the same as in the original one,
but we pay the price that the  proper distances are highly distorted.

The physical position of the event horizon is important due to the holographic
principle. The area of the horizon at a given time bounds the entropy of
the universe at this time by the projection of the inside degrees of freedom onto
the horizon by an holographic prescription; the way in which this projection is
made will give important insights on the nature of the way gravity must be consistent with
the quantum world; we hope our physical coordinates complement the conformal ones to
improve the understanding of the causal structure of the cosmological models.

We emphasize that our analysis has been made, in the main part of this work, using only the
kinematics inferred for the line element of the RW model and as a consequence our
results are valid for different dynamical scenarios; we require only the agreement
with the cosmological principle for the energy momentum tensor that couples to the geometry.
In this way we have shown that expanding universes with a final decelerating epoch
do not have event horizons. We have introduced the concept of asymptotic horizon
as the locus where the NGs change concavity and we have proved that for
an expanding cosmological model, if it is accelerating, its future horizon converges with
the asymptotic horizon. For universes with constant
concavity light rays the asymptotic horizon
matches the event horizon for all times; this is the case for the FRW quintessence dominated
universe and for the dS model.

The fact that we have an expression for the asymptotic distance of the event horizon can
be useful in view of the holographic conjecture; the event horizon is the natural place
where the degrees of freedom in the bulk are projected, and its area is a measure of
the entropy of the system; asymptotically the area of the horizon is then an absolute
upper bound on the number of degrees of freedom of the universe and is given
by the area of the asymptotic horizon.

We would like to comment also the teleological nature of the horizon. We cannot be sure at
a given time if a model has event horizons or not; this question depends on
the scale factor for all the times and specifically on the value the scale
factor takes asymptotically. What we can only do is to extrapolate our local
knowledge to all future times and extract the consequences. The horizon is a
fragile concept as that of the black hole; in this last case the horizon disappears as
time evolves but due to Hawking evaporation; in the cosmological case the \emph{probable}
horizon can be diluted but now due to kinematic arguments. Can this two processes be related?

To finish, let us remember that the origin of this research was motivated by the
conflict among the accelerating cosmologies and string theory; the observation
that the actual universe is accelerated and probably will continue this regime forever,
forbids the existence of physical states separated an infinite distance in an operational
sense; that is we cannot make measures on two states which distance is greater
than the horizon; however an S-matrix description requires asymptotic accessible states.
On the other hand we have been able to identify an accelerating model without event horizon.
We hope this work can help to resolve this challenge.

\section*{Appendix} \nonumber

In this Appendix we demonstrate the following statement:
\begin{quote}
\emph{The event horizon of any RW cosmological model,
regardless of the dynamical theory in which is based, converges to the asymptotic
horizon coinciding both at the future infinity.}
\end{quote}

We will show that asymptotically the intersection angle between
the NGs that traverse the asymptotic horizon and the proper asymptotic
horizon tend to zero.
In this way we can affirm that the asymptotic horizon tends to coincide with
a NG; this limiting NG will be the event horizon.
First we calculate the slope of the NG when it intersects the
asymptotic horizon;
introducing $D_{AsH}$ in (\ref{3-2}) and using (\ref{3-9}) and (\ref{3-10}):
\begin{equation} \label{5-1}
\dot{D}_{n}\mid _{AsH} \ =
\frac{D_{Hub}}{D_{AsH}}-1=
{-1-q \over q}=
\frac{\dot{D}_{Hub}}{1-\dot{D}_{Hub}}.
\end{equation}
Then we show that the previous expression for large times is equal to the slope
of the asymptotic horizon ($\dot{D}_{n}\mid _{AsH}=\dot{D}_{AsH}$). We do this
for the different values that takes the slope of the Hubble distance at infinite
time; we define $l \equiv  \lim_{t \rightarrow \infty} \dot{D}_{Hub}$. Because the universe
is accelerating $0 \leq l \leq 1$. We study the following cases.

\begin{description}
\item[A]
$l=0$ and $D_{Hub}\rightarrow H<\infty$ (\emph{Horizontal Asymptote}).

Using (\ref{5-1}) we have $\dot{D}_{n}\mid _{AsH} \rightarrow 0$;
then we have $q=-1$ at late times and
\begin{equation} \label{5-A}
D_{AsH}\rightarrow D_{Hub}\Rightarrow
D_{AsH}\rightarrow H \Rightarrow \dot{D}_{AsH}\rightarrow 0.
\end{equation}
Examples of this behavior are
finally dS dominated models (final exponential grow).
See Fig. \ref{figura3} left.

\item[B]
$l=0$ and $D_{Hub} \rightarrow \infty $ (\emph{Parabolic grow}).

Again $\dot{D}_{n}\mid _{AsH} \rightarrow 0$.
Now, because $D_{AsH}>D_{Hub}$, $D_{AsH} \rightarrow \infty $; so
together with (\ref{3-9}) and (\ref{3-10}) we use
the l'H\^opital theorem. We find a solution identical to the previous one
\begin{equation} \label{5-B}
\frac{D_{Hub}}{D_{AsH}}\rightarrow 1
\Rightarrow \frac{\dot{D}_{Hub}}{\dot{D}_{AsH}}\rightarrow 1
\Rightarrow \dot{D}_{AsH}\rightarrow 0.
\end{equation}
We don't know any simple realistic model obeying this behavior.

\item[C]
$l \in (0,1)$ (\emph{Oblique Asymptote}).

Also with (\ref{3-9}) and (\ref{3-10}) and the l'H\^opital theorem we find
\begin{equation} \label{5-C}
\lim_{t\rightarrow \infty}\frac{D_{AsH}}{D_{Hub}}=\frac{1}{1-l} \Rightarrow
\lim_{t\rightarrow \infty}\dot{D}_{AsH}=\frac{l}{1-l},
\end{equation}
which exactly corresponds to the limit of (\ref{5-1}). In this case $D_{Hub}$
and $D_{AsH}$ tend to a linear function. Examples of this behavior are finally
FRW quintessence dominated models (final potential grow).
See Fig. \ref{figura3} right.

\item[D]
$l=1$ (\emph{Limiting Oblique Asymptote}).

This case corresponds to an acceleration tending to zero from above
and it has been shown that there are not event horizons
(see subsection 3.3). With (\ref{5-C}) we find that now $\dot{D}_{AsH}$ diverges
and so $D_{AsH}$ has no asymptote.
Examples of this behaviour
are finally FRW models with $\omega \rightarrow -{1 \over 3}^{-}$ (final linear grow).
See Fig. \ref{figura4}.
\end{description}

We have covered all the possible finally accelerated scenarios and therefore
we have completed our demonstration.

The Table \ref{tabla2} resume our classification including the decelerated
universes (E and F) and an exotic accelerated universe (Sub A).

\begin{table}[!htb]
\begin{tabular}{|c|c|c|c|c|c|c|}
\hline
  & & & & & & \\
  Type & $\omega\rightarrow$ & $R(t) \sim$  & $D_{Hub}\rightarrow$ &
  $\dot{D}_{Hub}\rightarrow$ & $D_{AsH}\rightarrow$ & $D_{Hor}\rightarrow$\\
  & & & & & & \\
\hline \hline
 & & & & & & \\
   A &
   $-1^{+}$ &
   $e^{a t}$ &
   $\frac{1}{a}$&
   $0^{+}$ &
   $D_{Hub}$ &
   $D_{AsH}$ \\
 & & & & & & \\
 \hline
   & & & & & & \\
   B &
   $-1^{+}$ &
   $e^{a t^{n}}$ &
   $\frac{1}{n}\, t^{1-n}$&
   $0^{+}$ &
   $\frac{1}{n}\, t^{1-n}$ &
   $D_{AsH}$ \\
 & &$ ^{n<1}$& & & & \\
 \hline
   & & & & & & \\
   C &
   $(-1,-\frac{1}{3})$ &
   $t^{b}$ &
   $\frac{1}{b}\ t$&
   $\frac{1}{b}$ &
   $\frac{1}{b-1}\ t$ &
   $D_{AsH}$ \\
 & &$^{b>1}$ & & & & \\
  \hline
   & & & & & & \\
   D &
   $-\frac{1}{3}$ &
   $d \ t$ &
   $t$&
   $1$ &
   $\pm \infty$ &
   $\not\!\exists$ \\
 & & & & &$^{\dot{D}_{AsH}\rightarrow \infty}$ & \\
  \hline
   & & & & & & \\
   E &
   $(-\frac{1}{3},\ \infty)$ &
   $t^{b}$ &
   $\frac{1}{b}\ t$&
   $\frac{1}{b}$ &
   $-\frac{1}{1-b}\ t$ &
   $\not\!\exists$ \\
 & &$^{b<1}$ & & & & \\
  \hline \hline
   & & & & & & \\
   F &
   $+\infty$ &
   $\log \, t$ &
   $t \log \, t$&
   $\log \, t$ &
   $-t$ &
   $\not\!\exists$ \\
 & & & & & & \\
  \hline
   & & & & & & \\
   Sub A &
   $-1^{-}$ &
   $e^{a t^{n}}$ &
   $\frac{1}{n \, t^{n-1}}$ &
   $0^{-}$&
   $D_{Hub}$ &
   $D_{AsH}$ \\
 & &$ ^{n>1}$ & & &$^{D_{Hub}>D_{AsH}}$ & \\
 \hline
\end{tabular}
\caption{{\small Classification of the RW cosmological models based on the growing rate
of $D_{Hub}$ associated with different ranges for $\omega$.
The values for $D_{AsH}$ and $D_{Hor}$ are also exposed.
Compare each case with its graphic representation. The last two rows corresponds
to pathological universes.}}
\label{tabla2}
\end{table}

Finally we want to comment an interesting property.
It can be shown that if the event horizon has an asymptote, then it is the unique
NG having an asymptote. The rest of the NGs reconverge or recede without asymptote.
Using the l'H\^opital heorem in the NG equation (\ref{3-2}) we found that
if $\lim \dot{D}_{n}\neq \infty$ then
\begin{equation} \label{5-2}
\lim_{t\rightarrow\infty} \dot{D}_{n}=
\lim_{t\rightarrow\infty}\frac{ \dot{D}_{n}}{ \dot{D}_{Hub}}-1 \Rightarrow
\lim_{t\rightarrow\infty} \dot{D}_{n}=\frac{l}{1-l}\in (0,\infty)
\end{equation}
which is identical to (\ref{5-C}) and can be easily extrapolated
to (\ref{5-A}). So, for the cases A and C we can prescribe this rule:
\emph{the event horizon is the unique NG having an asymptote and
the asymptotic horizon has exactly the same asymptote}.
This property could be used for a better delimitation of the event horizons by
means of the Legendre's transformation which relates a mathematical function to
its family of tangent straight lines.

\section*{Acknowledgements} \nonumber
We have benefit for useful discussions with L. Bell and I. Shapiro.
This work was supported by MCYT (Spain), grant FPA2000-1252.


\begin{thebibliography}{21}

\bibitem{r1}
S.~Hellerman, N.~Kaloper and L.~Susskind,
``String theory and quintessence,''
JHEP {\bf 0106} (2001) 003
[arXiv:hep-th/0104180].

\bibitem{r2}
W.~Fischler, A.~Kashani-Poor, R.~McNees and S.~Paban,
``The acceleration of the universe, a challenge for string theory,''
JHEP {\bf 0107} (2001) 003
[arXiv:hep-th/0104181].

\bibitem{Maldacena}
J.~Maldacena,
``The large $N$ limit of superconformal field theories and supergravity,''
Adv.\ Theor.\ Math.\ Phys.\  {\bf 2} (1998) 231
[Int.\ J.\ Theor.\ Phys.\  {\bf 38} (1998) 1113]
[arXiv:hep-th/9711200].

\bibitem{r4}
A.~Strominger,
``The dS/CFT correspondence,''
JHEP {\bf 0110} (2001) 034
[arXiv:hep-th/0106113].

\bibitem{r5}
M.~Li,
``Matrix model for de Sitter,''
[arXiv:hep-th/0106184].

\bibitem{'tHooft}
G.~'t Hooft,
``Dimensional reduction in quantum gravity,''
[arXiv:gr-qc/9310026].

\bibitem{susskind}
L.~Susskind,
``The World as a hologram,''
J.\ Math.\ Phys.\  {\bf 36} (1995) 6377
[arXiv:hep-th/9409089].

\bibitem{Perlmutter}
S.~Perlmutter {\it et al.}  [Supernova Cosmology Project Collaboration],
``Measurements of Omega and Lambda from 42 High-Redshift Supernovae,''
Astrophys.\ J.\  {\bf 517} (1999) 565
[arXiv:astro-ph/9812133].

\bibitem{Ellis}
G.~F.~R.~Ellis and T.~Rothman,
``Lost horizons,''
Am.\ J.\ Phys.\  {\bf 61} (1993) 883.


\bibitem{velocidades}
T.~M.~Davis and C.~H.~Lineweaver,
``Superluminal Recession Velocities,''
Cosmology and Particle Physics 2000
Conference Proceedings
[arXiv:astro-ph/0011070].


\bibitem{Ellis2}
M.~Rauch,
``Comments on `Lost Horizons' by G. F. R. Ellis and T. Rothman [Am. J. Phys. 61
(10), 883-893 (1993)],''
Am.\ J.\ Phys.\  {\bf 63} (1995) 87;
G.~F.~R.~Ellis,
``Past light cone shape and refocusing in cosmology, A Response to Michael Rauch's
`Comments on Lost Horizons' [Am. J. Phys. 63, 87 (1995)],''
Am.\ J.\ Phys.\  {\bf 63} (1995) 88.

\bibitem{Fishlersusskind}
W.~Fischler and L.~Susskind,
``Holography and cosmology,''
[arXiv:hep-th/9806039].


\bibitem{bousso}
R.~Bousso,
``A Covariant Entropy Conjecture,''
JHEP {\bf 9907} (1999) 004
[arXiv:hep-th/9905177].

\bibitem{HE}
S.~W.~Hawking and G.~F.~R.~Ellis,
``The large-scale structure of space time,''
Cambridge University Press, Cambridge (1973).

\bibitem{encadenada}
T.~M.~Davis, C.~H.~Lineweaver and J.~K.~Webb,
``Solutions to the chained galaxy problem and the
observation of receding blue-shifted objects,''
[arXiv:astro-ph/0104349].





\end{thebibliography}
\end{document}